\documentclass[journal]{IEEEtran}

\usepackage [dvips]{graphicx}
\usepackage{amsmath}
\usepackage{amssymb}
\usepackage{graphics,epsf}
\usepackage{float}
\usepackage{color}
\usepackage{bm}
\usepackage{subfigure}
\usepackage{wrapfig}
\usepackage{algorithm,algorithmic}
\usepackage[bookmarks=yes]{hyperref}
\usepackage{flushend}
\usepackage{pgf,tikz}
\usepackage{epstopdf}

\begin{document}
	\title{Adaptive Video Streaming over LTE Unlicensed\thanks{The authors are with the Department of Electrical and Computer Engineering, University of Thessaly, Greece.}}
	
\author{Apostolis Galanopoulos and Antonios Argyriou~\IEEEmembership{Senior Member,~IEEE}\vspace{-6mm}
\IEEEcompsocitemizethanks{
\IEEEcompsocthanksitem
}}

\markboth{Submitted to IEEE Transactions on Multimedia, \today }{Submitted to IEEE Transactions on Multimedia, \today}

	\maketitle

\begin{abstract}	
In this paper we consider the problem of adaptive video streaming over an LTE-A network that utilizes Licensed Assisted Access (LAA) which is an instance of LTE Unlicensed. With LTE Unlicensed users of the LTE-A network opportunistically access radio resources from unlicensed and licensed carriers through Carrier Aggregation (CA). Our objective is to select the highest possible video quality for each LTE-A user while also try to deliver video data in time for playback, and thus avoid buffer under-run events that deteriorate viewing experience. However, the unpredictable nature of the wireless channel, as well as the unknown utilization of the unlicensed carrier by other unlicensed users, result in a challenging optimization problem. We first focus on developing an accurate system model of the adaptive streaming system, LTE-A, and the stochastic availability of unlicensed resources. Then, the formulated problem is solved in two stages: First, we calculate a proportionally fair video quality for each user, and second we execute resource allocation on a shorter time scale compatible with LTE-A. We compare our framework with the typical proportional fair scheduler as well as a state-of-the-art LTE-A adaptive video streaming framework in terms of average segment quality and number of buffer under-run events. Results show that the proposed quality selection and scheduling algorithms, not only achieve higher video segment quality in most cases, but also minimize the amount and duration of video freezes as a result of buffer under-run events.
\end{abstract}


\section{Introduction}
The performance of cellular networks today is limited primarily by the available wireless spectrum~\cite{ericsson-2015}. As mobile devices are becoming capable of running applications that demand a considerable amount of bandwidth, e.g. high quality video streaming, new challenges rise for the Mobile Network Operators (MNO). As the user needs increase, the 3rd Generation Partnership Project (3GPP) intends keep up with it by introducing several enhancements as part of the next releases of LTE Advanced (LTE-A).

One major enhancement is Carrier Aggregation (CA) that has already been part of Release 10 of the standard and is used to aggregate up to 5 Component Carriers (CC) and thus increase the data rate available to the users~\cite{CA_qualcomm}. A number of band combinations for aggregation as well as several types of CA have been proposed ever since but the major problem of spectrum scarcity still remains a challenge. A promising solution that is employed by MNOs is Spectrum Refarming (SR), through which under-utilized spectrum reserved for old Radio Access Technologies (RATs) is reassigned to LTE-A. The number of legacy devices that utilize the aforementioned spectrum decreases as they migrate to the newer technology, e.g. LTE-A, so a portion of it can be redistributed to this new technology \cite{SR}. This technique however, requires that refarmed spectrum belongs to the same MNO.

As a result 3GPP has proposed Licensed Assisted Access (LAA) \cite{3gpp_laa} to exploit Unlicensed Bands (UB) in LTE-A systems. LAA is a realization of the more general concept of LTE Unlicensed which entails the exploitation of unlicensed spectrum by LTE systems. The enabling technology behind LAA is CA. However, LAA allows CA to occur between a component carrier that belongs to the MNO Licensed Band (LB) and possibly others that belong to an UB that is potentially occupied by other users. Hence, the additional spectrum provided by UBs may not always be exploitable for LTE-A communication since LAA must ensure that UB users continue to utilize the UB spectrum (almost) unaffected~\cite{CA_qualcomm}.

In this paper we suggest that video streaming is an application that can experience significant performance boost with the adoption of LAA in current LTE-A systems, and also with the use of both licensed and unlicensed resources in future 5G systems. Adaptive video streaming protocols such as Dynamic Adaptive Streaming over HTTP (DASH) \cite{DASH} try to deliver video data to mobile users by measuring the average throughput of the wireless channel and delivering the video file in segments of a quality level that is less or equal to this throughput. This is due to the fact that the higher the video quality, the higher the encoding rate of the video. Consequently, the required data rate that the user must achieve in order to watch a video with high quality is increased. The actual throughput of the communication channel can vary significantly over time due to the unstable nature of the wireless channel, making its estimation challenging. Consequently, a DASH client typically tries to fill the video playback buffer when the channel is good in order to cope with unexpected bad channel at a later point in time. In addition to the channel quality, the unpredictable availability of unlicensed resources makes the problem of adaptive video streaming even more challenging.



\subsection{Carrier Aggregation \& Licensed Assisted Access in LTE-A} \label{LAA}
\begin{figure}
	\begin{center}
		\includegraphics[width=\columnwidth]{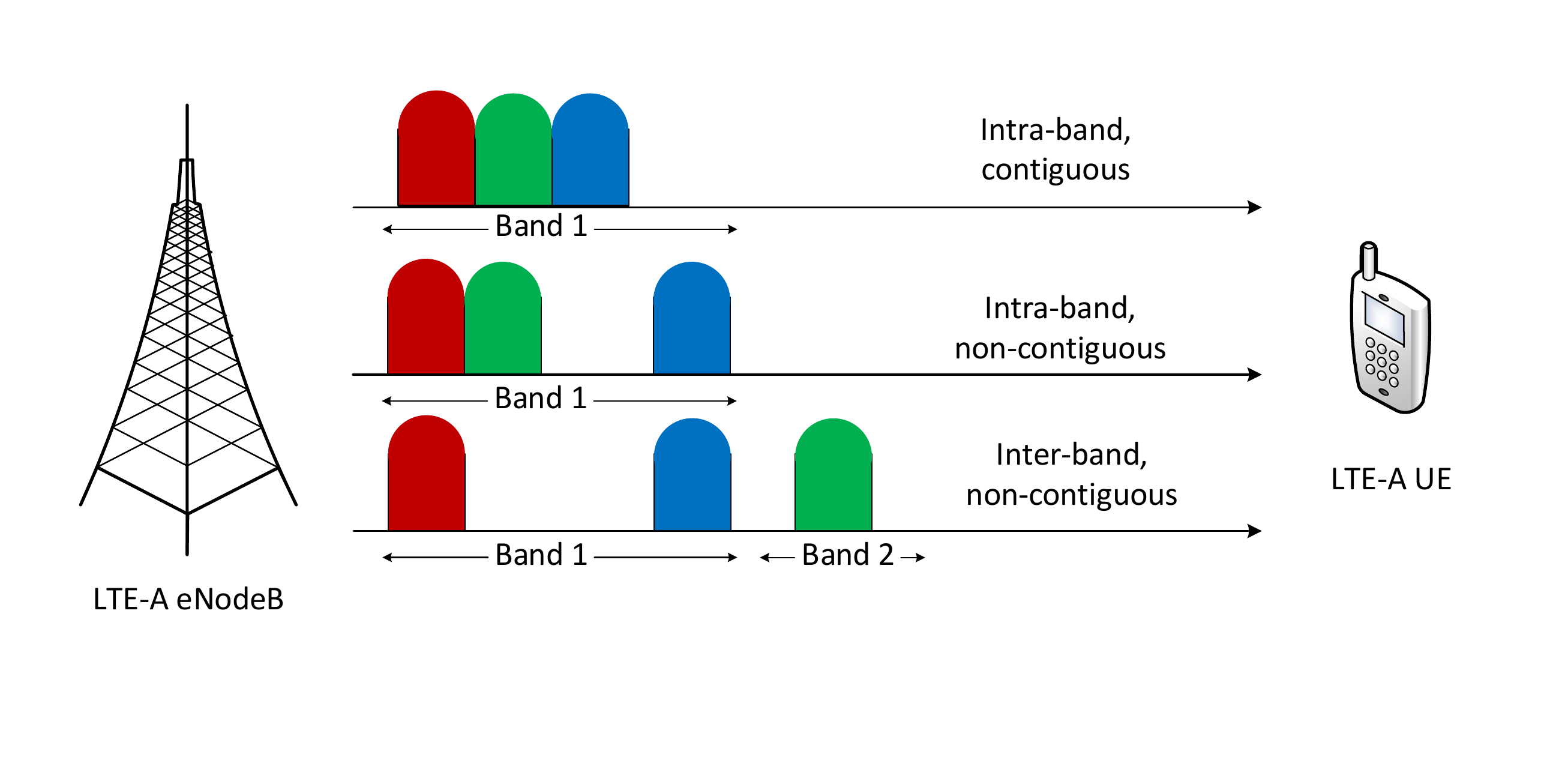}
		\caption {Forms of Carrier Aggregation.}
		\label{CA_types}
	\end{center}
\end{figure}

Carrier Aggregation is a technology used in LTE-A to increase the available bandwidth and thus achieve the target data rates set for 4G cellular communications. LTE supports the following bandwidths per CC: 1.4, 3, 5, 10, 15, 20 MHz \cite{3gpp_lte}. Several CCs of possibly different bandwidths can be used by an eNodeB to allocate resources on multiple CCs, provided that the User Equipment (UE) is CA-enabled and can decode a CA signal. Three types of CA exist depending on the availability of carriers that determine the physical layer architecture of the communicating pair. In intra-band contiguous CA all aggregated CCs belong to the same frequency band and occupy contiguous carrier center frequencies (Figure~\ref{CA_types}). In intra-band non-contiguous CA all CCs belong to the same frequency band, however not all of them employ contiguous carrier positions wherein the band. Finally, in inter-band non-contiguous CA, the CCs belong to different bands and thus are not contiguous in frequency.

3GPP has introduced the concept of utilizing UB through CA to improve users' data rate with LAA~\cite{3gpp_laa}. A LAA system employs at least 2 CCs one of which is in a LB and the rest of CCs are in an UB where other systems may operate. This requires an intra-band non-contiguous CA implementation at the eNodeB since the aggregated CCs belong to different bands. The 5 GHz band is mainly considered for LAA due to the significant amount of available spectrum but there are several restrictions in the utilization of the band in order to avoid interfering with other systems. Furthermore, each country has defined different regulations regarding the utilization of the sub-bands that constitute the 5 GHz band. To this end, and since 3GPP aims in a global application of LAA, the frequency chunk that is expected to be utilized by LAA systems is supposed to be accessed mainly by WiFi users. LAA systems need to incorporate a series of functionalities that will ensure the smooth operation of such systems. These main functionalities defined in \cite{3gpp_laa}, have been studied in \cite{laa_functionalities} and are summarized next:
\begin{itemize}
	\item Listen Before Talk: Perform a Clear Channel Assessment (CCA) prior to LTE transmission to ensure an idle channel that will not interfere to other systems' transmission.
	\item Carrier Selection: The aggregated CC in the UB should be on a low traffic/interference condition concerning the activity of other systems.
	\item Discontinuous Transmission: LAA transmissions cannot occupy the UB indefinitely, and give the chance to other systems that compete for the same channel to transmit their data.
	\item Transmit Power Control. Regulations in different regions of the world impose a maximum transmit power level in unlicensed bands.
\end{itemize}
It is clear that all the above requirements make challenging the access to the UB spectrum. A LAA system may schedule its users in multiple CCs through CA but whether the UB CC will be available or not, highly depends on the activity of other systems and the aforementioned functionalities.

\subsection{Related work and motivation}
Video delivery in LTE-based cellular networks has been extensively studied in recent years. In \cite{green_video} the energy efficient delivery of DASH video segments is studied in a LTE heterogeneous network. The problems of user association and resource allocation are studied jointly so that users can download video files with the highest quality possible, but also minimize the total power dedicated for radio transmission and network backhaul. The authors in \cite{joint_TD_resource_partitioning} propose a time domain resource partitioning technique for video streaming in heterogeneous networks. The NP-hard problem that is formulated, jointly optimizes resource/rate allocation as well as the selected video quality and is then decomposed into simpler problems using a primal-dual approximation algorithm. In \cite{adaptive_video_whitespace} the concept of adaptive video streaming with Scalable Video Coding (SVC) over a shared frequency band is studied. The dynamics of the unlicensed users are modeled by a Markov decision process and are incorporated to the system in order to make optimal decisions about the quality of the future segment requests. In \cite{RA_joint_CA}, \cite{app_aware_RBA} the problem of resource allocation in LTE CA systems is considered. The solution involves rate allocation of multiple CCs to the UEs of the network by maximizing logarithmic and sigmoidal like utility functions that represent user satisfaction. In addition, a distributed version of the resource allocation algorithm is presented, that is based on UE bidding for resources process. In \cite{AVIS} a scheduling framework (AVIS) for adaptive video streaming over cellular networks is presented. The authors propose a gateway level architecture for AVIS by implementing it in two entities. The first one is responsible for deciding the encoding rate for each user, while the second one allocates resources in a way that the users' average data rate is kept stable so that segments are downloaded on time. While this framework is in many ways similar to the one proposed in this work, it lacks exploitation of UE buffer status as well as unlicensed band availability information. In \cite{stochastic_scheduling} resource allocation is achieved by an interference mitigation scheme for heterogeneous networks. A stochastic scheduling algorithm is applied to schedule resources probabilistically, that is also observed to increase femtocell capacity. The works in \cite{Neely_1}, \cite{Neely_2} study admission/congestion control and transmission scheduling in small cell networks for adaptive video streaming. More specifically in \cite{Neely_1}, a network utility maximization problem is formulated in order to keep transmission queues of helper nodes stable. The admission control policy problem is tackled by choosing the helper node as the one with the smallest queue backlog. Transmission scheduling requires the maximization of sum rates with the queue backlogs serving as weights. Furthermore in \cite{Neely_2}, an algorithm is proposed that calculates the pre-buffering and re-buffering time for each user so that they can experience a smooth streaming service without buffer under-run events. A similar work is also presented in \cite{Neely_helpers}, where users are able to download from a number of base stations and decide which of them is better to serve them.

Even though adaptive video streaming frameworks have been extensively studied before, to the best of our knowledge our work is the first that proposes a detailed model and a comprehensive optimization framework for the application of adaptive video streaming over LAA. The applicability of our scheme is not limited to LAA: 5G systems will make extensive use of unlicensed spectrum to increase data rate and serve demanding applications like video streaming. Hence, modeling and optimization tools must be developed to this aim.

\section{System model}
\begin{figure}[t]
	\begin{center}
		\includegraphics[keepaspectratio,width = 0.9\linewidth]{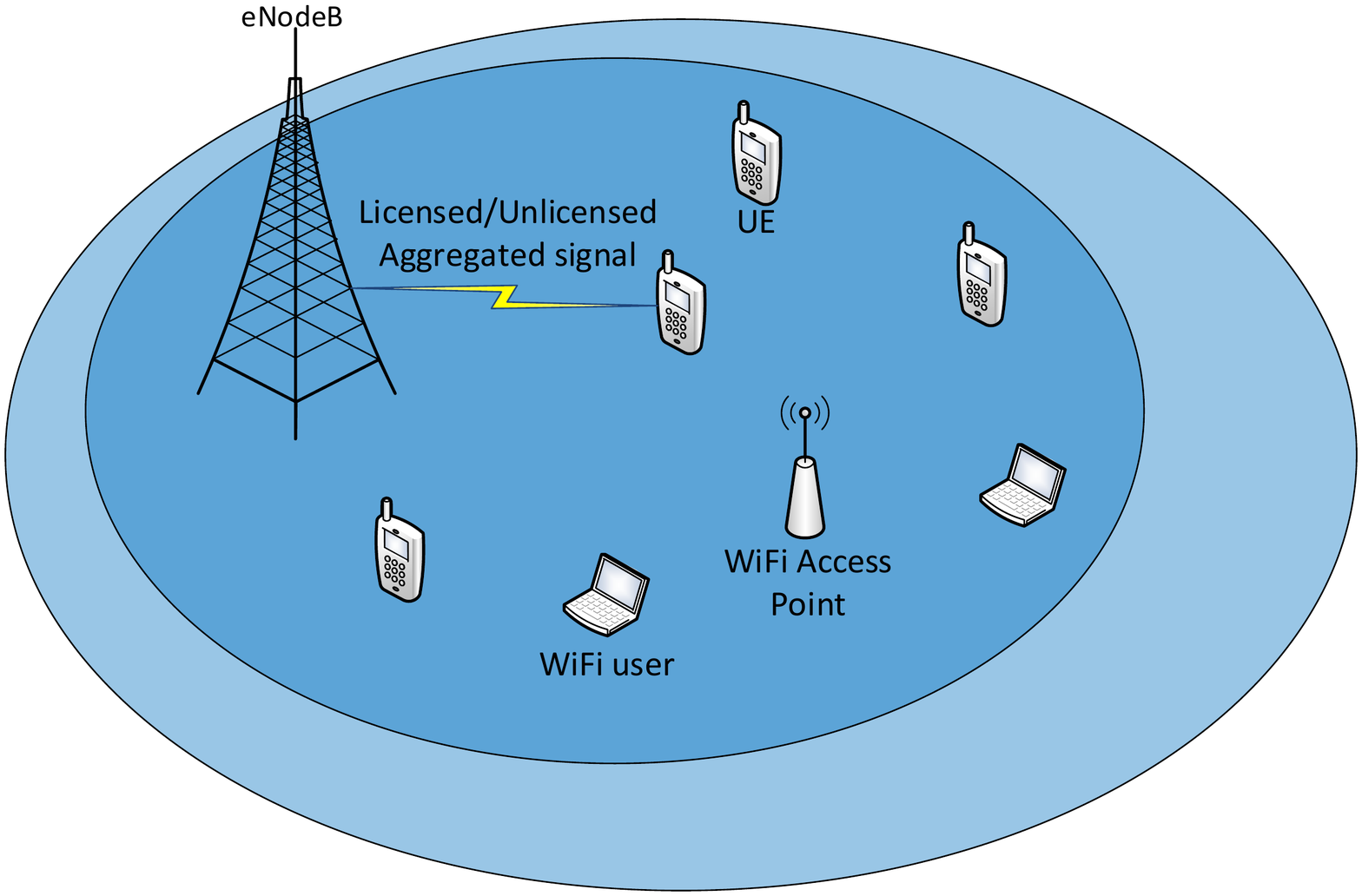}
		\caption{The considered network topology consists of LTE-A UEs that access both licensed and unlicensed spectrum, while the WiFi users use only the unlicensed spectrum.}
		\label{System model}
	\end{center}
\end{figure}

We consider a LTE-A eNodeB with LAA capabilities, i.e. the functionalities described in Section \ref{LAA}, enabling it to monitor traffic in one or more unlicensed band CCs. During each scheduling interval $t$, the eNodeB can employ CA to schedule resources from a licensed primary CC and an unlicensed secondary CC, each one of bandwidth $W_L$ and $W_U$ respectively. Depending on the values of $W_L$ and $W_U$ a number of resource blocks (RBs) $M_L$ and $M_U$ are available for scheduling. Each RB consists of 12 sub-carriers spaced at 15 KHz providing a total bandwidth of $W=180\ KHz$ per RB. A set of UEs $\mathcal{K}$ exists in the area of the eNodeB and each user $k \in \mathcal{K}$ requests DASH video files. The network model is depicted in Figure \ref{System model}. UEs are in the coverage area of the licensed carrier (light blue color) and the unlicensed one (dark blue color), where WiFi systems also operate. The coverage area of the unlicensed carrier is typically smaller because it is centered at a higher frequency.

\subsection{Video streaming}
The encoding rate of each video chunk is decided by the UE approximately every 10 seconds (each chunk no matter its size corresponds to 10 seconds video duration). We denote this time period as the Quality Selection Interval (QSI). The eNodeB determines the video quality for each UE depending on channel quality, secondary CC availability, and playback buffer status of the UEs. Since the QSI is relatively long, it is reasonable to assume that each UE can communicate all necessary information such as buffer status and channel state to the eNodeB through the uplink channel. Channel quality and secondary CC availability are averages observed by the UEs and the eNodeB respectively. The buffer status is the remaining video playback time that is already stored to each UE's playback buffer. We denote the set of available encoding rate levels for the video file requested by UE $k$ as~\cite{joint_TD_resource_partitioning}:

\begin{equation}
	\mathcal{D}^k = \{D^k_1,D^k_2,...,D^k_L\}
\end{equation}
where $D^k_i$ is the encoding rate of the video quality $i$ in bits per second. Furthermore, we assume that the encoding rate (i.e. video quality) increases with the index $i$, so that $D^k_i > D^k_{i-1}$. $L$ is the number of encoding rates in which the requested video is available by the server. An example of recommended encoding rates for different quality levels is provided in Table \ref{encoding_rates}, where the frame rate is 30 frames per second and the aspect ratio, i.e. the ratio of row to column pixels, is 16:9 that are typical values.

\begin{table}
	\begin{center}
		\begin{tabular} {| c | c | c |}
			\hline
			\textbf{Quality} & \textbf{Resolution} & \textbf{Bitrate (Kbps)} \\ \hline
			360p & 640$\times$360 & 1000 \\ \hline
			480p & 848$\times$480 & 2500 \\ \hline
			720p & 1280$\times$720 & 5000 \\ \hline
			1080p & 1920$\times$1080 & 8000 \\ \hline
			1440p & 2560$\times$1440 & 10000 \\ \hline
			2160p & 3840$\times$2160 & 35000 \\ \hline
		\end{tabular}
	\end{center}
	\caption{Video quality levels and encoding rates.}
	\label{encoding_rates}
\end{table}

\subsection{Unlicensed band traffic estimation}

The eNodeB is required to monitor the activity of the UB in order to coexist harmoniously with the deployed UB systems as proposed by 3GPPP \cite{3gpp_laa}. This requires the existence of an energy detector that is capable of collecting samples periodically, and determining whether a signal is present in the desired channel. The statistics of the energy detector can be used to estimate the probability that the channel is idle, and thus an LTE transmission can take place.
The energy detector collects samples of the unlicensed band and determines whether a transmission takes place or not based on an energy threshold. This energy threshold is typically at -82 dBm for the Carrier Sensing mechanism of WiFi. In \cite{coexistence} however, it is stated that the energy detection level that is used for LAA is set higher at -62 dBm possibly interrupting WiFi transmissions. Thus, the activity of WiFi stations in the unlicensed band can be estimated by measuring the number of samples resulting in busy medium versus the total number of samples collected over a specified time period. So, by defining $N_1$ to be the number of samples where the detected energy was above the energy threshold, and as $N_2$ the respective value for below the energy threshold detections we have:

\begin{equation}
	P_{on} = \frac{N_1}{N_1+N_2}
\end{equation}
being the probability that the unlicensed band is occupied and

\begin{equation}
	\label{P_off_1}
	P_{off} = \frac{N_2}{N_1+N_2}
\end{equation}
the probability that the unlicensed band is idle at some random time instance.

The works of Bianchi \cite{Bianchi_1},~\cite{Bianchi_2} provide a solid mathematical framework for modeling the channel access probability of WiFi users using a discrete time Markov chain are considered, in order to calculate $P_{off}$ under several realistic scenarios. Particularly in~\cite{Bianchi_2}, the probability that a WiFi station transmits at a random slot is given as:
\begin{equation}
	\label{wifi_tr}
	\gamma = \frac{2(1-2p)}{(1-2p)(W_{in}+1)+pW_{in}(1-(2p)^i)}
\end{equation}
where $W_{in}$ is the minimum backoff window of 802.11 Distributed Coordination Function (DCF), $p$ is the probability that a transmitted packet collides and $i$ is the maximum number of times the backoff window is doubled after consecutive packet collisions. Assuming a number of $n$ stations want to transmit a packet during a slot, $p$ is given by
\begin{equation}
	\label{wifi_col}
	p = 1-(1-\gamma)^{n-1},
\end{equation}
since at least one of the remaining $n-1$ stations should also transmit so that a collision occurs. By solving \eqref{wifi_tr} and \eqref{wifi_col} we obtain the values of $\gamma$ and $p$. The probability $p_{tx}$ that during a random WiFi slot there is at least one transmission that can be detected by the LAA eNodeB (perfect detection is assumed) is given by:
\begin{equation}
	p_{tx} = 1-(1-\gamma)^n
\end{equation}
We define $\Psi$ to be the random variable that represents the number of consecutive idle slots between two WiFi transmissions. Then the average $\Psi$ is given by:
\begin{equation}
	\mathop{\mathbb{E}}\{\Psi\} = \frac{1}{p_{tx}}-1
\end{equation}
where $\mathop{\mathbb{E}}\{\cdot\}$ denotes the expectation of a random variable. One more thing is required for the calculation of $P_{off}$. That is the average duration of a packet transmission in WiFi slots, since $\mathop{\mathbb{E}}\{\Psi\}$ is also calculated in WiFi slots. Assuming that this value is known as $\mathop{\mathbb{E}}\{P\}$ then $P_{off}$ is calculated as:

\begin{equation}
	\label{P_off_2}
	P_{off} = \frac{\mathop{\mathbb{E}}\{\Psi\}}{\mathop{\mathbb{E}}\{\Psi\} + \mathop{\mathbb{E}}\{P\}}
\end{equation}
$P_{off}$ is therefore a function of $W_{in}$, $i$, $n$ and $\mathop{\mathbb{E}}\{P\}$. From these parameters only $n$ and $\mathop{\mathbb{E}}\{P\}$ can vary during the system operation and can therefore affect the performance of our LAA system. Of course these values are unknown to eNodeB which in practice will perform energy detection based on \eqref{P_off_1}. This model however is required to simulate realistic WiFi traffic scenarios for performance evaluation.

\subsection{Solution approach}
\begin{figure}[t]
	\begin{center}
		\includegraphics[width=\columnwidth]{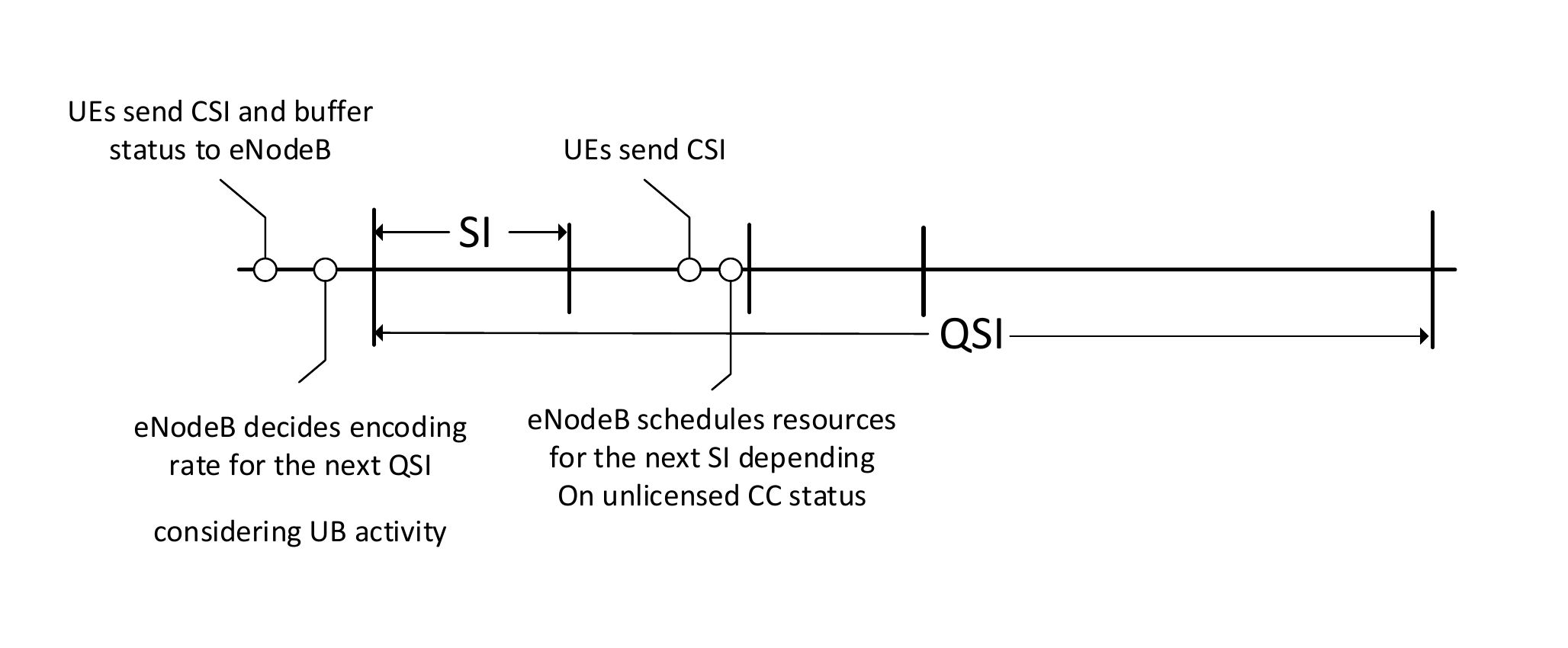}
		\caption{Quality selection and resource allocation decision timeline.}
		\label{Time graph}
	\end{center}
\end{figure}

After quality selection decisions are made, resource allocation is handled by eNodeB every 10 milliseconds, i.e. the duration of one LTE frame. RBs are scheduled to the UEs depending on their recent Channel State Information (CSI) as well as the current availability of the secondary CC. We denote this 10 ms resource allocation duration as Scheduling Interval (SI). Note that each QSI interval consists of 1000 SIs. For further clarity we illustrate the network actions versus time in Figure~\ref{Time graph}. In the following two sections the problems of quality selection and resource allocation are analyzed under the described setting.

\section{Quality Selection}
For eNodeB to decide the video quality of the chunks to be delivered for the next QSI, assuming that the current QSI is $T$, certain information is required. Each UE $k \in \mathcal{K}$ reports their average Signal to Noise Ratio (SNR) experienced of QSI $T$. These values are denoted by $SNR_L^k(T)$ and $SNR_U^k(T)$ for the licensed and unlicensed CCs respectively. In addition, the UEs report the status of their playback buffer.

\subsection{Modeling Buffer Dynamics}
\begin{figure}
	\begin{center}
		\includegraphics[width=\columnwidth]{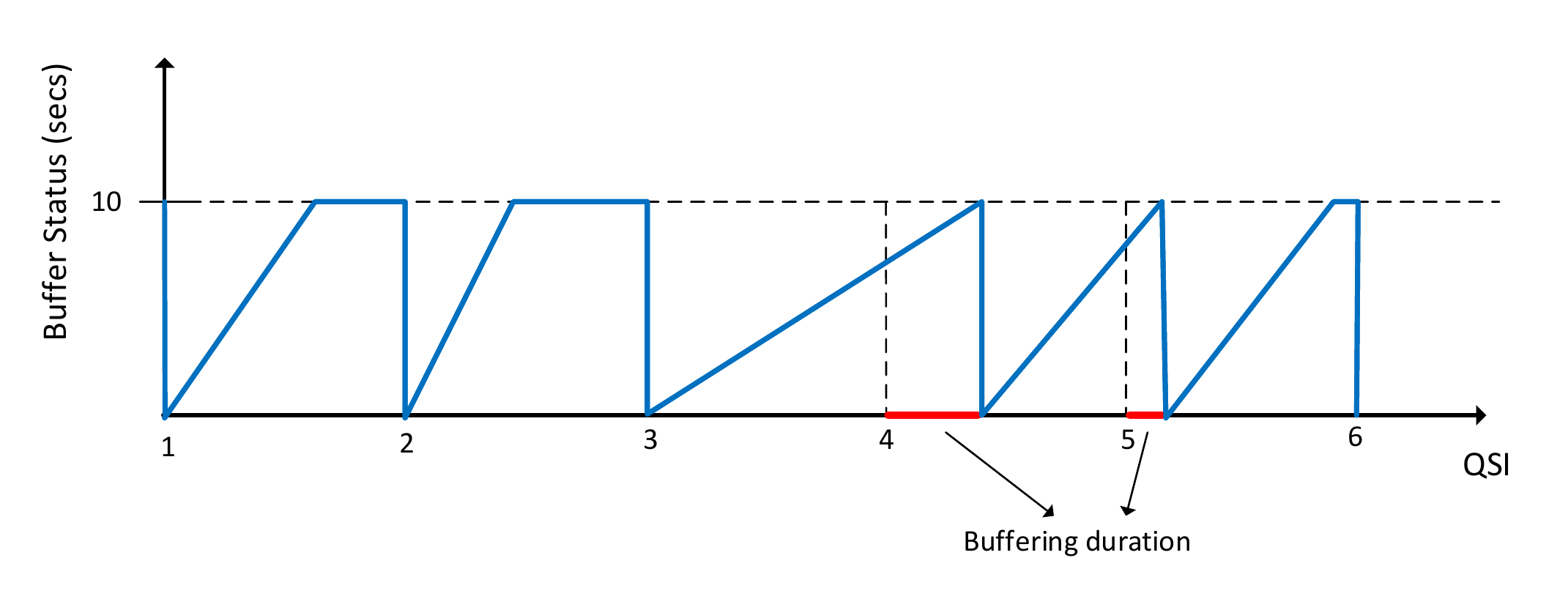}
		\caption{Modeling buffer dynamics.}
		\label{Buffer status}
	\end{center}
\end{figure}

During QSI $T$ a UE downloads the segment to be displayed during the next QSI $T+1$. The duration of the buffered video during QSI $T$ for UE $k$ is denoted as $B^k(T)$ and is updated for the next QSI as follows:
%
%
\begin{equation} \label{Buffer equation}
	B^k(T+1) =
	\begin{cases}
		\frac{S^k(T)}{R^k(T)}, & \text{if } B^k(T)=10 \\
		B^k(T)+\frac{S^k(T)}{R^k(T)}, & \text{if } B^k(T)+\frac{S^k(T)}{R^k(T)} < 10 \\
		B^k(T)+\frac{S^k(T)}{R^k(T)}-10, & \text{if } B^k(T)+\frac{S^k(T)}{R^k(T)} \geq 10
	\end{cases}
\end{equation}
$B^k(T)$ is the video duration in seconds stored at QSI $T$ and $\frac{S^k(T)}{R^k(T)}$ is the video duration downloaded at QSI $T$. $S^k(T)$ denotes the size of the segment(s) in bits that will be delivered during QSI $T$, while $R^k(T)$ denotes the average download rate during QSI $T$. We also assume that at the beginning of each QSI $T$, the buffer contents are reduced by an amount equal to 10 seconds worth of playable video if the video segment of the previous QSI $T-1$ has been downloaded. If this is not the case, the buffer occupancy at the beginning of QSI $T$ is $\frac{S^k(T)}{R^k(T)}$, which is lower than 10 seconds and the completion of the segment download will occur at some point during the next QSI. An example is presented in Figure \ref{Buffer status} and explained in the next paragraphs, helps to illustrate these concepts further.

Let us assume that at the beginning of QSI 1 the buffer contains the first 10-second segment which is downloaded before playback starts. Immediately it is delivered to the application layer and the downloading of the next segment begins filling the buffer again. The slope of the buffer status shows the rate at which the segment is downloaded. For the first two QSIs everything runs smoothly and segments are downloaded on time. This is captured by the first leg of equation \eqref{Buffer equation} where for example $B^k(2)=\frac{S^k(2)}{R^k(2)}=10$ and thus $B^k(3)=10$. However, at the third QSI the segment to be delivered to the application at QSI 4 has not been downloaded until QSI 4 begins so that we have  $B^k(4)=\frac{S^k(3)}{R^k(3)}<10$. Since $B^k(4) < 10$, in order to calculate $B^k(5)$ we use the second or third leg of equation \eqref{Buffer equation} depending on whether the segment was finally downloaded and delivered to the application during QSI 4. Indeed, since $B^k(4)+\frac{S^k(4)}{R^k(4)} > 10$ we have that the previous segment was downloaded at some point during QSI 4 and the buffer is updated as: $B^k(5)=B^k(4)+\frac{S^k(4)}{R^k(4)}-10$. The period from the beginning of QSI 4 until the segment is delivered and the buffer empties, is the buffering duration when the application layer buffer awaits a segment delivery and the video freezes. The same situation occurs during QSI 5 but with shorter buffering duration. Finally, there is one last case for the next QSI buffer status update that is not captured in Figure \ref{Buffer status} and that is the second leg of equation \eqref{Buffer equation}. In this case, the segment that had not been downloaded at some QSI $T$, was still not received by the end of the next QSI $T+1$ (i.e. $B^k(T)+\frac{S^k(T)}{R^k(T)}<10$). If this happens the buffering duration lasts for the entire QSI $T$. However, this case is extremely rare, since additional weight will be given to the scheduling of users not able to download their segments fast enough, as will be shown in the next section.

\subsection{Utility maximization for video quality selection }
The estimated availability of the unlicensed CC is based on the activity of WiFi users, as sensed by eNodeB, possibly by energy detection~\cite{laa_functionalities},~\cite{coexistence}. The probability that the unlicensed CC is idle for QSI $T$ is denoted by $P_{off}$ and is re-evaluated periodically by eNodeB. The download rate of UE $k \in \mathcal{K}$ that benefits from both licensed and unlicensed CCs at QSI $T$ is then given by:
\begin{equation}
	\label{total_rate}
	R^k(T) = R_L^k(T) + P_{off} R_U^k(T)
\end{equation}
To calculate $R^k(T)$, one has to perform resource allocation in both licensed and unlicensed CCs by accounting the UEs' SNR values and buffer status, as well as the expected availability of the unlicensed CC. The resulting $R^k(T)$ can serve as an upper bound on the encoding rate of the chunk to be delivered to user $k$ at QSI $T$. The data rates gained by the licensed and unlicensed carriers are defined as
\begin{equation}
	R_L^k(T) = x_L^k M_L W \log(1+SNR_L^k(T)),
\end{equation}
and
\begin{equation}
	\label{unlicensed_rate}
	R_U^k(T) = x_U^k M_U W \log(1+SNR_U^k(T)),
\end{equation}
where $x_L^k(T)$ and $x_U^k(T)$ represent the fraction of the $M_L$ and $M_U$ sets of RBs to be allocated to UE $k$ from the licensed and unlicensed carriers respectively. Therefore we have
\begin{equation}
	x_L^k(T) \in [0,1],\ \forall k \in \mathcal{K},
\end{equation}
and
\begin{equation}
	x_U^k(T) \in [0,1],\ \forall k \in \mathcal{K}.
\end{equation}
Thus, we define the utility function for each UE $k$ based on $R^k(T)$ and $B^k(T)$, in order to decide the resource allocation that will select the quality for QSI $T$ as follows:
\begin{equation} \label{utility}
	U^k(T) = \log (R^k(T) + \alpha B^k(T))
\end{equation}
In the above $\alpha$ is a biasing factor that will affect the impact of the UEs' buffer status on resource allocation and therefore on quality selection decisions. Notice how the buffer status affects rate allocation: Due to the logarithmic function, UEs with reduced buffer contents will tend to be allocated more resources so that they can download more data and avoid a buffer underflow. Equation \eqref{utility} is a concave function with respect to $x_L^k(T)$ and $x_U^k(T)$ so we can formulate the sum utility maximization problem $\bm{P_1}$ as:
\begin{equation} \label{utility_max}
	\bm{P_1}: \max_{\bm{x_L},\bm{x_U}} \sum_{k \in \mathcal{K}} U^k(T)
\end{equation}
Subject to:
\begin{equation} \label{C_1}
	\sum_{k \in \mathcal{K}} x_L^k(T) = 1
\end{equation}
\begin{equation} \label{C_2}
	\sum_{k \in \mathcal{K}} x_U^k(T) = 1
\end{equation}
Vector $\bm{x_L}$ contains variables $x_L^k(T),\ \forall k \in \mathcal{K}$, while vector $\bm{x_U}$ contains variables $x_U^k(T),\ \forall k \in \mathcal{K}$.

This problem is convex because of the logarithmic function and the linear constraints. To solve it we define the augmented Lagrangian function by embedding the constraints (\ref{C_1}) and (\ref{C_2}) in the objective function:
\begin{equation}
	\begin{split}
		L(\bm{x_L},\bm{x_U},\lambda,\mu) = I_k U^k(T) - \lambda (I_k x_L^k(T) - 1) - \mu (I_k x_U^k(T) - 1) \\
        -\frac{\rho}{2} (I_k x_L^k(T) - 1)^2 - \frac{\rho}{2} (I_k x_U^k(T) - 1)^2	
	\end{split}
\end{equation}
where $\lambda$, $\mu$ are Lagrangian multipliers for each of the two constraints, $I_k$ is a unitary row vector of length $|\mathcal{K}|$ and $\rho > 0$.

To maximize $L$ we perform the Alternating Direction Method of Multipliers (ADMM) \cite{ADMM}, which involves optimizing $L$ over each variable separately at each iteration $\tau$. Formally we have:
\begin{equation} \label{ADDM_1}
	\bm{x_L}^{\tau+1} := \arg\max_{\bm{x_L}} L(\bm{x_L},\bm{x_U}^{\tau},\lambda^{\tau},\mu^{\tau})
\end{equation}
\begin{equation} \label{ADMM_2}
	\bm{x_U}^{\tau+1} := \arg\max_{\bm{x_U}} L(\bm{x_L}^{\tau+1},\bm{x_U},\lambda^{\tau},\mu^{\tau})
\end{equation}
\begin{equation} \label{ADDM_3}
	\lambda^{\tau+1} := \lambda^{\tau} + \rho (I_k \bm{x_L}^{\tau+1}-1) + \rho (I_k \bm{x_U}^{\tau+1}-1)
\end{equation}
\begin{equation} \label{ADMM_4}
	\mu^{\tau+1} := \mu^{\tau} + \rho (I_k \bm{x_L}^{\tau+1}-1) + \rho (I_k \bm{x_U}^{\tau+1}-1)
\end{equation}
Each of the equations (\ref{ADDM_1}) - (\ref{ADMM_2}) is solved by setting the partial derivative of $L$ equal to zero, and solving for each variable, which is then used to obtain the Lagrangian variables for iteration $\tau+1$ through (\ref{ADDM_3}) - (\ref{ADMM_4}). When ADMM converges, the optimal solution of $\bm{P_1}$ is found and the achievable data rate of each UE can be calculated. This upper bound of data rate will determine the quality of the video segment to be delivered to the UE as the maximum available from $\mathcal{D}^k$ which does not exceed the achievable data rate $R^k(T)$. The proposed Quality Selection Algorithm is described in Algorithm \ref{Algo_1}.

\begin{algorithm}[t]
	\caption{Quality Selection}
	\label{Algo_1}
	\begin{algorithmic}
		\FOR {\textbf{each} QSI $T$}
			\STATE{\textbf{Require:} $SNR_L^k(T), SNR_U^k(T), B^k(T),\ \forall k \in \mathcal{K}$, $P_{off}$}
			\STATE{Initialize $\bm{x_L}^0, \bm{x_U}^0, \lambda^0, \mu^0$}
			\STATE{$\tau \leftarrow 0$}
			\REPEAT
				\STATE{Calculate $\bm{x_L}^{\tau+1}, \bm{x_U}^{\tau+1}, \lambda^{\tau+1}, \mu^{\tau+1}$ by eq. (\ref{ADDM_1})-(\ref{ADMM_4})}
				\STATE{$\tau \leftarrow \tau+1$}
			\UNTIL{Optimal solution found}
			\FOR {\textbf{each} $k \in \mathcal{K}$}
				\STATE{Calculate $R^k(T)$ by eq. \eqref{total_rate}-\eqref{unlicensed_rate}}
				\STATE{$D^k(T) \leftarrow \max \mathcal{D}^k, \{D^k(T) \leq R^k(T)\}$}
			\ENDFOR
		\ENDFOR
	\end{algorithmic}
\end{algorithm}

\begin{figure}
	\begin{center}
		\includegraphics[width=\columnwidth]{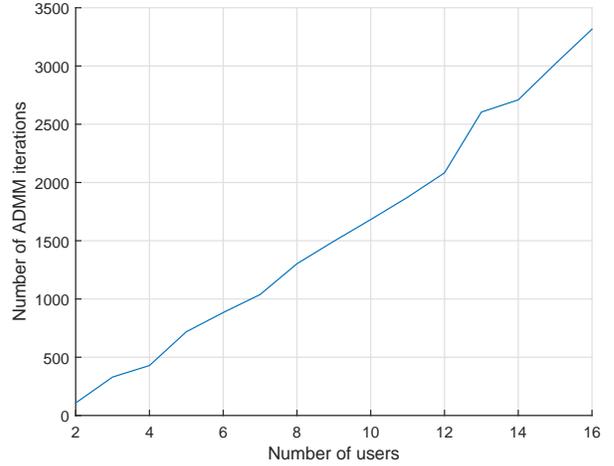}
		\caption{Average ADMM iterations versus number of users.}
		\label{ADMM_iterations}
	\end{center}
\end{figure}

The test for convergence is carried out as follows \cite{ADMM}. If $\bm{x_L}^*$, $\bm{x_U}^*$, $\lambda^*$ and $\mu^*$ are the optimal values obtained from the ADMM algorithm, the following conditions must be satisfied:
\begin{equation} \label{optimality_1}
	\frac{\partial L(\bm{x_L}^*,\bm{x_U}^*,\lambda^*,\mu^*)}{\partial \bm{x_L}} = 0
\end{equation}
\begin{equation} \label{optimality_2}
	\frac{\partial L(\bm{x_L}^*,\bm{x_U}^*,\lambda^*,\mu^*)}{\partial \bm{x_U}} = 0
\end{equation}
\begin{equation} \label{violation_1}
	I_k \bm{x_L}^* = 1
\end{equation}
\begin{equation} \label{violation_2}
	I_k \bm{x_U}^* = 1
\end{equation}
Conditions \eqref{optimality_1} and \eqref{optimality_2} ensure that the solution is optimal, while \eqref{violation_1} and \eqref{violation_2} that it does not violate the problem's constraints. Proof of the linear convergence of ADMM is provided in \cite{ADMM_convergence}. Figure \ref{ADMM_iterations} displays the average number of iterations required for ADMM to converge to the optimal solution versus the number of users ($K$) that clearly affect the number of problem variables and thus the convergence rate of ADMM. It is evident that the increase of iteration steps is linearly proportional to $K$, which makes the application of Algorithm \ref{Algo_1} an efficient solution to problem $\bm{P_1}$.

\section{Resource Block scheduling}
In this section we propose a RB scheduling policy \cite{scheduling_policies} suitable for both licensed and unlicensed CCs. To deal with resource allocation on a frame basis, a stochastic optimization framework is required \cite{Neely_book}. The reason is that each 10-second QSI, consists of 1000 10-ms SIs during which the scheduler must ensure that all segments to be transmitted to the UEs at the current QSI, are received by the UEs before the end of the QSI. Unfortunately, no information about future channel states and UB availability is known to eNodeB at each SI and thus a stochastic optimization approach is necessary.

\subsection{Problem formulation}
At the beginning of each QSI $T$ eNodeB has retrieved the video segments of a specific quality from the video server according to the solution proposed in the previous section. A number of $K$ queues hold the segments (one queue per user), and eNodeB must schedule RBs to UEs during each SI with the goal of emptying all queues by the end of the current QSI. This implies that the segments of QSI $T$ are delivered in time to the UEs so that they can watch freeze-free video content. The number of bits stored at the queues of the eNodeB at each SI $t$ is denoted by $Q^k(t)$. For the first SI the queues are initialized with the segment size of the current QSI $T$ as
\begin{equation}
	Q^k(t=1) = S^k(T),\ \forall k \in \mathcal{K},
\end{equation}
and are updated for the next SI by
\begin{equation}
	Q^k(t+1) = Q^k(t) - \frac{r^k(t)}{100},\ \forall k \in \mathcal{K},
\end{equation}
where $r^k(t)$ is the data rate in bits per second experienced by UE $k$ at SI $t$.

The unlicensed carrier can either be available or unavailable at each SI according to the UB activity. The auxiliary variable $a_u(t)$ indicates if the unlicensed carrier is available at SI $t$ and thus, unlicensed RBs can be allocated at a specific SI:
\begin{equation}
	a_u(t)=
	\begin{cases}
		1, & \text{with probability}\ P_{off} \\
		0, & \text{with probability}\ P_{on}
	\end{cases}
\end{equation}
The SNR experienced by each UE $k \in \mathcal{K}$ at each SI $t$ is denoted by $SNR_L^k(t)$ and $SNR_U^k(t)$ for the licensed and unlicensed band respectively. The maximum throughput a UE can achieve at the licensed and unlicensed carriers is calculated using the current SNR. We introduce the scheduling variable $y_{mk}(t)$ which is defined as:
\begin{equation}
	y_{mk}(t)=
	\begin{cases}
		1, & \text{if RB } m \text{ is allocated to user}\ k \\
		0, & \text{otherwise}
	\end{cases}
\end{equation}
The data rate of UE $k$ at each carrier is then calculated as
\begin{equation}
	r_L^k(t) = \sum_{m \in \mathcal{M}_L} y_{mk}(t) W \log (1 + SNR_L^k(t))
\end{equation}
for the licensed carrier and
\begin{equation}
	r_U^k(t) = \sum_{m \in \mathcal{M}_U} y_{mk}(t) W \log (1 + SNR_U^k(t))
\end{equation}
for the unlicensed carrier. The total achievable data rate of user $k$ at SI $t$ is then calculated as:
\begin{equation}
	r^k(t) = r_L^k(t) + a_u(t) r_U^k(t)
\end{equation}

The goal is to schedule resources at each SI $t$ so that by the end of the QSI $T$, each UE $k$ will have downloaded the $S^k(T)$ bits of the video chunk. The problem is formally expressed as:
\begin{equation}
	\label{max_rate}
	\max_{y_{mk(t)}} \sum_{k \in \mathcal{K}} \sum_{t=1}^{1000} r^k(t).
\end{equation}
subject to:

\begin{equation}
	\sum_{t=1}^{1000} \frac{r^k(t)}{100} \geq S^k(T),\ \forall k \in \mathcal{K}
\end{equation}
where $\frac{r^k(t)}{100}$ denotes the amount of bits downloaded by user $k$ during SI $t$, since each SI lasts for 10 milliseconds and $r^k(t)$ is given in bits per second.

\subsection{Backlog and Channel Aware Scheduling Policy}
The calculation of $y_{mk}(t) \ \forall t \in [1,...,1000]$ in \eqref{max_rate} requires prior knowledge of $SNR_L^k(t)$, $SNR_U^k(t)$ and $a_u(t) \forall t \in [1,...,1000]$, which is unavailable at the start of QSI $T$. To tackle this problem we propose a scheduling policy for each SI $t$. It accounts for the current channel conditions and unlicensed CC availability: $SNR_L^k(t), SNR_U^k(t), a_u(t)$ and the current backlog of the user $Q^k(t)$. The proposed algorithm is based on the \textit{max-weight} algorithm \cite{Neely_book}, where scheduling decisions are made based on current queue backlogs and channel states without the need of knowing the channel.

There are several scheduling policies for LTE systems in the literature both for time and frequency domain scheduling~\cite{scheduling_policies}. Leveraging the Orthogonal Frequency Division Multiple Access (OFDMA) technology of LTE we allocate LTE RBs to different UEs at each SI $t$, applying thus frequency domain scheduling. Under this type of scheduling, a metric function $\delta(m,k)$ is calculated and then each RB $m$ is iteratively allocated to the UE for which the metric function obtains the highest value. Proportional Fair Scheduling (PFS) for example, considers the users instant data rate on each RB as well as the average rate experienced by each user in order to formulate the metric function. In our case however, each UE reports one SNR value for each CC, $SNR_L^k(t), SNR_U^k(t)$ respectively. It is possible though to require sub-band instead of wide-band level feedback reports and thus acquire feedback in the form of $SNR_L^k(m,t), SNR_U^k(m,t)$, where $SNR_L^k(m,t)$ is the SNR experienced by UE $k$ on SI $t$ for RB $m$ of the licensed CC and $SNR_U^k(m,t)$ is the respective value for the unlicensed CC. Whichever the case, $r^k(t)$ can be decomposed to a series of data rates given by the different reported SNRs, whether they differ per RB or they are the same for RBs of the same CC as
\begin{equation}
	 r^k(t) = \sum_{m \in \mathcal{M}_L} y_{mk}(t) r_L^k(m,t)  + a_u(t) \sum_{m \in \mathcal{M}_U} y_{mk}(t) r_U^k(m,t),
\end{equation}
where $r_L^k(m,t)$ is given by
\begin{equation} \label{rate_per_RB_L}
	r_L^k(m,t) = W \log(1 + SNR_L^k(m,t)),
\end{equation}
and $r_U^k(m,t)$ by
\begin{equation} \label{rate_per_RB_U}
	r_U^k(m,t) = W \log(1 + SNR_U^k(m,t))
\end{equation}
The last two equations are the data rates in bits per second experienced by UE $k$ on SI $t$ and on RB $m$ if $m$ belongs to the licensed and unlicensed bands respectively. In our system we assume a wide-band feedback reporting system and thus equations \eqref{rate_per_RB_L} and \eqref{rate_per_RB_U} reduce to
\begin{equation}
	r_L^k(m,t) = W \log(1 + SNR_L^k(t))
\end{equation}
and
\begin{equation}
	r_U^k(m,t) = W \log(1 + SNR_U^k(t))
\end{equation}
After defining the users' instant data rate per RB we proceed by calculating the average throughput experienced by UE $k$ until SI $t$ as:
\begin{equation}
	\overline{r^k}(t) = \frac{\sum_{n=1}^{t} r^k(n)}{t}.
\end{equation}
With standard PFS \cite{scheduling_policies} the metric function $\delta(m,k)$ is calculated as:
\begin{equation} \label{metric_1}
	\delta(m,k) =
	\begin{cases}
		\frac{r_L^k(m,t)}{\overline{r^k}(t)}, & \text{if } m \in \mathcal{M}_L \\
		\frac{r_U^k(m,t)}{\overline{r^k}(t)}, & \text{if } m \in \mathcal{M}_U
	\end{cases}
\end{equation}
It is evident from the form of $\delta(m,k)$ that it is maximized for UEs that experience high instant data rate for RB $m$ and low average data rate, providing thus the proportional fairness characteristic of the metric function. Equation \eqref{metric_1} is calculated iteratively for each RB and each UE and RB $m$ is allocated to the optimal UE $k^*$ that maximizes $\delta(m,k)$. Formally $k^*$ is given as:
\begin{equation}
	k^* = \arg\max_k \delta(m,k),\ \forall m \in \mathcal{M}_L \cup \mathcal{M}_U
\end{equation}


In our system, proportional fairness is implemented with the logarithmic function in the objective function of the utility maximization problem $\bm{P_1}$. At this stage, the proposed scheduling policy empties all backlog queues by the end of each QSI. Since each QSI consists of 1000 SIs the desired scheduling policy must result in: $Q^k(1000) = 0,\ \forall k \in \mathcal{K}$. Inspired by the \textit{max-weight} algorithm, the proposed scheduling metric is obtained by multiplying the data rate experienced by each UE for the current SI, with by user queue backlog. The scheduling decision is based on which UE maximizes the newly defined metric:
\begin{equation} \label{metric_2}
	\delta(m,k) =
	\begin{cases}
		Q^k(t) r_L^k(m,t), & \text{if } m \in \mathcal{M}_L \\
		Q^k(t) r_U^k(m,t), & \text{if } m \in \mathcal{M}_U
	\end{cases}
\end{equation}
%
%
Each RB is iteratively assigned to the UE that maximizes \eqref{metric_2}. Its respective queue backlog is updated and the procedure continues until all RBs are allocated. The proposed Backlog and Channel Aware Scheduling Policy (BCASP) is presented in Algorithm \ref{Algo_2}.

\begin{algorithm}[t]
	\caption{Backlog and Channel Aware Scheduling Policy}
	\label{Algo_2}
	\begin{algorithmic}
		\FOR {\textbf{each} SI $t$}
			\STATE{\textbf{Require:} $SNR_L^k(t), SNR_U^k(t), S^k(T), Q^k(t),\ \forall k \in \mathcal{K}$, $a_u(t)$}
			\FOR {\textbf{each} $m \in \mathcal{M}_L$}
				\FOR {\textbf{each} $k \in \mathcal{K}$}
					\STATE {$\delta(m,k) \leftarrow Q^k(t) r_L^k(m,t)$}
				\ENDFOR
				\STATE{$k^* \leftarrow \arg\max_k \delta(m,k)$}
				\STATE{$Q^{k^*}(t) \leftarrow Q^{k^*}(t)- \frac{r_L^{k^*}(m,t)}{100}$}
			\ENDFOR	
			
			\IF {$a_u(t) = 1$}
				\FOR {\textbf{each} $m \in \mathcal{M}_U$}
					\FOR {\textbf{each} $k \in \mathcal{K}$}	
						\STATE {$\delta(m,k) \leftarrow Q^k(t) r_U^k(m,t)$}
					\ENDFOR
					\STATE{$k^* \leftarrow \arg\max_k \delta(m,k)$}
					\STATE{$Q^{k^*}(t) \leftarrow Q^{k^*}(t)- \frac{r_U^{k^*}(m,t)}{100}$}
				\ENDFOR	
			\ENDIF	
		\ENDFOR
	\end{algorithmic}
\end{algorithm}

\section{Performance evaluation}
In this section extensive simulation results on the considered LAA video streaming framework are provided. In the first two sub-sections, the simulation setup is explained both in link level (LTE physical layer) as well as in system level, in order to lay the foundation for the upcoming simulations and the explanation of results in the following sub-section.

\subsection{Link level simulation setup}
\begin{figure}
	\begin{center}
		\includegraphics[width=\columnwidth]{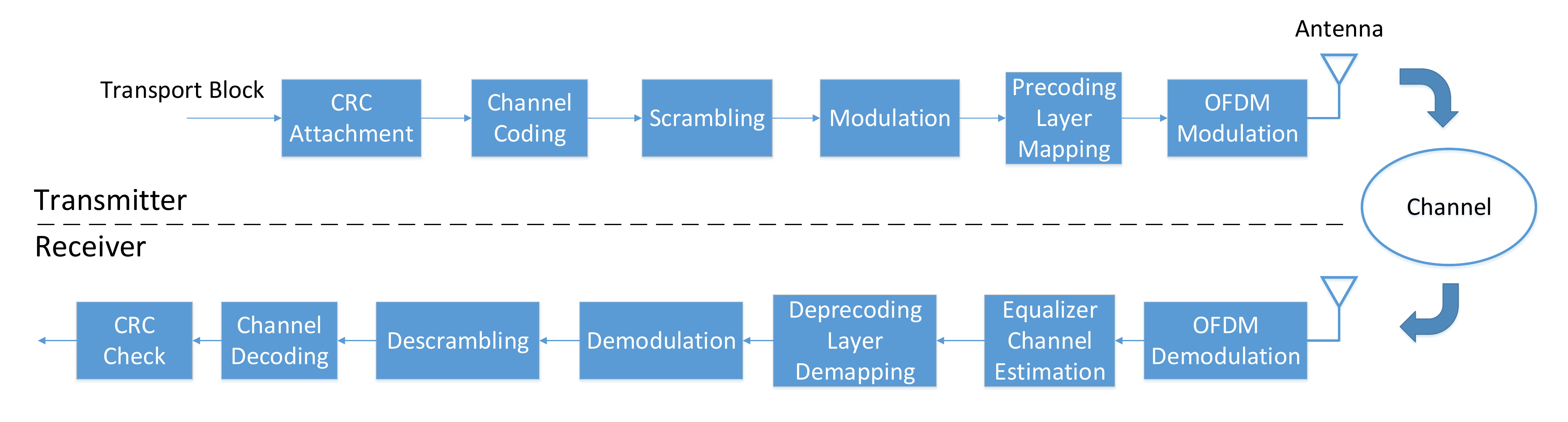}
		\caption{LTE Physical Layer downlink processing chain.}
		\label{LTE_chain}
	\end{center}
\end{figure}

The link level simulation consists of the implementation of LTE Physical Layer in MATLAB in order to obtain performance metrics of the downlink transmission of LTE under different channel conditions and transmission modes. The LTE downlink processing chain is provided in Figure \ref{LTE_chain}. The functionality of the blocks depicted in Figure \ref{LTE_chain} is briefly described as follows:

\begin{itemize}
	\item CRC Attachment. Cyclic Redundancy Check (CRC) is used for error detection of the transport blocks. A number of parity bits are attached at the end of the transport block.
	
	\item Channel Coding. Turbo coding with a rate of 1/3 is employed to protect the transmission of data against channel fading. 
	
	\item Scrambling. The input bit streams are combined to scrambling sequences in order to produce pseudo-random codewords.
	
	\item Modulation. The scrambled bits are used to generate complex valued modulation symbols. The constellations supported by LTE are QPSK, 16QAM and 64QAM in order to provide adaptive modulation.
	
	\item Precoding and Layer Mapping. This functionality is employed in case of a multi-antenna transmission scheme. It involves the mapping of the input symbols on each layer for transmission on the available antenna ports.
	
	\item OFDM Modulation. Orthogonal Frequency Division Multiplexing modulation is applied by utilizing Inverse Fast Fourier Transform (IFFT) in order to convert a frequency selective fading channel to a number of flat fading orthogonal sub-channels of narrower bandwidth.
\end{itemize}

The reverse actions are made at the receiver side with the addition of channel estimation which is used to counteract the effects of the wireless channel on the received signal and improve Bit Error Rate (BER). The procedure involves the reception of known pilot symbols in specific Resource Elements of the LTE Resource Grid. The receiver can then estimate the effect of the channel by observing the difference between the known and received pilot symbols.

The simulation environment developed to model the physical layer described above sets system parameters such as the number of frames, bandwidth, transmission scheme etc and passes the OFDM modulated signals through a fading channel for a series of SNR values. The receiver decodes the received signals and the throughput performance of the link is calculated.

\begin{figure}
	\begin{center}
		\includegraphics[width=\columnwidth]{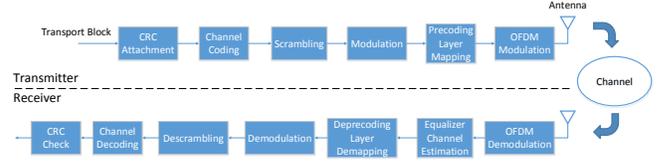}
		\caption{Physical layer throughput versus SNR.}
		\label{LTE_PHY_sim}
	\end{center}
\end{figure}

Throughput results of the simulator are presented in Figure \ref{LTE_PHY_sim}. The duration of the simulations was 100 LTE frames and the channel model used was the Extended Pedestrian A model. Adaptive modulation was employed according to \cite{CQI_mapping}. The figure depicts all supported transmission schemes for different bandwidth and Tx/Rx antenna setups. In the following sections a specific setup will be used for all links in the cellular network in order to provide system level results of the UE scheduling for a video streaming application.

\subsection{System level simulation setup}
Now we describe the details of the LAA system that provides streaming services with the functionalities described in the previous sections. We consider a cell topology with a number of $K$ UEs spread uniformly in a $2 \times 2$ kilometer area and the LAA enabled eNodeB at the center of it. One licensed and one unlicensed CCs of 20 MHz, each one entailing a number of $M_L=M_U=100$ RBs are considered. The link level profile of both CCs is assumed to be $4\times1$ transmit diversity as displayed in Figure \ref{LTE_PHY_sim}.


The SNR in dB that each UE $k$ experiences for the licensed CC at each QSI $T$ is given by:
\begin{equation} \label{snr_l}
	SNR^k_L(T) = P^{rx}_L(dBm) - N_0(dBm)
\end{equation}
where $P^{rx}_L(dBm)$ is the received power in dBm for the licensed CC, and $N_0(dBm)$ is the noise power. In order to calculate $P^{rx}_L(dBm)$, large scale fading with path loss and log-normal shadowing is considered as:
\begin{equation}
	P^{rx}_L(dBm) = P^{tx}(dBm) - FSPL_L(dB) - X_s
\end{equation}
where $P^{tx}(dBm)$ is the transmit power of eNodeB in dBm, $FSPL_L(dB)$ is the Free Space Path Loss (FSPL) for the licensed CC in dB and $X_s$, which accounts for log-normal shadowing, is a Gaussian random variable with $\mathcal{N}(0,\sigma^2)$. $FSPL_L(dB)$ is given by
\begin{equation}
	FSPL_L(dB) = 20 (\log_{10}(d) + \log_{10}(f_L) - 7.378),
\end{equation}
where $d$ is the distance between transmitter and receiver in meters, and $f_L$ is the center frequency of the transmitted signal in the licensed band in Hertz. The unlicensed CC belongs to a different spectrum band and thus FSPL is given by 
\begin{equation}
    FSPL_U(dB) = 20 (\log_{10}(d) + \log_{10}(f_U) - 7.378),
\end{equation}
where $f_U$ is the center frequency of the unlicensed CC. Thus, for the unlicensed CC we get that:
\begin{equation}
	P^{rx}_U(dBm) = P^{tx}(dBm) - FSPL_U(dB) - X_s
\end{equation}
and
\begin{equation} \label{snr_u}
	SNR^k_U(T) = P^{rx}_U(dBm) - N_0(dBm).
\end{equation}
The average SNR experienced during QSI $T-1$ is then used to calculate $SNR^k_L(T)$ and $SNR^k_U(T)$ as:
\begin{equation}
	SNR^k_L(T) = \frac{\sum_{t=1}^{1000} SNR^k_L(t)}{1000}
\end{equation}
and
\begin{equation}
	SNR^k_U(T) = \frac{\sum_{t=1}^{1000} SNR^k_U(t)}{1000}
\end{equation}
respectively.

The SNRs calculated above consider only large scale fading which is constant during the segment downloading.To capture small scale fading which occurs even with minor movements of the receiver, we incorporate Rayleigh fading that changes the experienced SNR from one SI to the next. The details of the values used for the parameters of the system level simulation setup are provided in Table \ref{system parameters}.

\begin{table}
	\begin{center}
		\begin{tabular} {| l | c |}
			\hline
			\textbf{Parameter} & \textbf{Value} \\ \hline
			$W_L/W_U$ & 20 MHz \\ \hline
			$M_L/M_U$ & 100 \\ \hline
			$f_L/f_U$ & 2.1/5.8 GHz \\ \hline
			$P^{tx}$ & 43 dBm \\ \hline
			$N_0$ & -80 dBm \\ \hline
			$\sigma^2$ & 3 \\ \hline
		\end{tabular}
	\end{center}
	\caption{System level simulation setup parameters.}
	\label{system parameters}
\end{table}

Concerning the video files that the UEs request for downloading, we consider a video file encoded in 6 different quality levels as described in Table \ref{encoding_rates}. Formally we have that:
\begin{equation} \label{segment_qualities}
	\mathcal{D}^k = \{1000, 2500, 5000, 8000, 10000, 35000\} Kbps,\ \forall k \in \mathcal{K}
\end{equation}
For the WiFi system setup we assume that a random number of $n$ WiFi stations are involved in packet transmissions during each QSI. Each packet is considered to have a fixed size of 1.5 KB and can be transmitted at a set of physical data rates supported by 802.11n that is operational in the 5GHz spectrum. This set $\mathcal{R}_w$ of physical WiFi data rates is as follows:
\begin{equation}
	\mathcal{R}_w = \{7.2,14.4,21.7,28.9,43.3,57.8,65,72.2\}\ Mbps
\end{equation}
The rate which will be selected comes from WiFi's rate adaptation mechanism and depends on each station's channel conditions. For a fixed packet size one can calculate the transmission duration for each possible data rate and by accounting that each WiFi slot lasts for 9 $\mu$s the set of possible transmission durations in number of WiFi slots is given by:
\begin{equation}
	\label{wifi_slots}
	\mathcal{T}_w = \{186,94,62,47,32,24,22,19\}\ \text{slots}
\end{equation}
Assuming that in each WiFi slot there is a number of $n$ WiFi stations that want to transmit a packet, we can calculate $P_{off}$ for each QSI by equations \eqref{wifi_tr}-\eqref{P_off_2} and \eqref{wifi_slots}. However, since the number of competing WiFi stations during a QSI is variable, we control $P_{off}$ so that its effect is evident, that is we assume that it follows a Gaussian distribution over QSIs as $\mathcal{G}(\mu,\phi^2)$.

\subsection{Simulation results}
For the remainder of this section, the performance of the proposed quality selection and scheduling algorithms will be tested through simulations performed using the setup described above. These simulations aim to highlight the effect of several network parameters such as the number of users and the unlicensed band availability, on crucial performance metrics such as average segment quality and number of video freezes. The proposed solution that consists of Algorithms \ref{Algo_1} and \ref{Algo_2}, both implemented on the eNodeB are compared to standard PFS and the AVIS framework \cite{AVIS}.

With PFS, the eNodeB allocates resources according to the scheduling metric in \eqref{metric_1}. Each UE experiences an average data rate, according to which it requests the appropriate quality for the next segment using the same rule as in Algorithm \ref{Algo_1}, i.e. it requests the maximum available segment encoding rate that does not exceed the average experienced data rate. Since PFS is a general solution and is not specifically designed to address the complex problem of adaptive video streaming, the AVIS framework is also employed for comparison. AVIS consists of two entities. The \textit{allocator}, which considers the resource requirements of the UEs and decides the encoding rate of the segments to be delivered to each UE, and the \textit{enforcer} which allocates resources in a similar manner to PFS so that UEs can download the desired segments in time. 

\subsubsection{Video Segment Quality}
\begin{figure}
	\begin{center}
		\includegraphics[keepaspectratio,width = 0.9\linewidth]{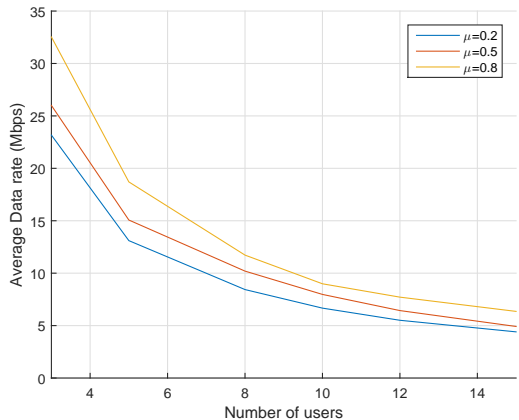}
		\caption{Average data rate vs number of UEs for different cases of unlicensed CC availability.}
		\label{data_rate}
	\end{center}
\end{figure}

The chosen segment quality depends on the data rate that the network can provide to each UE. Furthermore, data rate is a function of the number of UEs associated with the eNodeB as well as the number of available resources which directly links to unlicensed CC availability. To highlight the impact of the above parameters on average data rate, Figure \ref{data_rate} is provided.

Unlicensed CC availability is affected by the mean value $\mu$ of $P_{off}$ which varies in Figure \ref{data_rate}, while its standard deviation remains constant at $\phi^2=0.1$. As the number of UEs $K$ increases the average data rate decreases since more UEs share the same number of resources. The effect of unlicensed CC utilization can be seen for the three cases of $\mu$ in Figure \ref{data_rate}. For higher values of $\mu$ the unlicensed CC access probability is higher, UEs can be allocated with more resources and thus enjoy higher data rates that can lead to better video quality and/or fewer video freezes. The increased data rate effect is stronger when the number of UEs is small and deteriorates as $K$ increases. However, between different cases of $\mu$ the percentile drop in average data rate remains steady. For example, the average data rate is increased by approximately 40\% from $\mu=0.2$ to $\mu=0.8$ for all values of $K$. This indicates that even when $K$ is high and the average data rate is relatively small, the increased performance due to unlicensed CC utilization is far from negligible.

\begin{figure}
	\begin{center}
		\includegraphics[keepaspectratio,width = 0.9\linewidth]{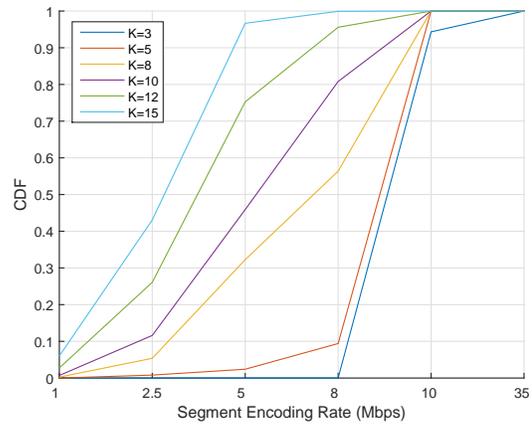}
		\caption{Segment quality CDF for different number of UEs.}
		\label{quality_cdf}
	\end{center}
\end{figure}

As stated before, the average segment quality mainly depends on the number of UEs, since the more UEs are active in the network, the fewer resources are allocated to each one, and thus they generally experience lower throughput which leads the designed quality selection algorithm to choose lower quality segments. Figure \ref{quality_cdf} displays the Cumulative Distribution Function (CDF) of the different segment qualities of \eqref{segment_qualities} for different number of UEs $K$ and for a number of 100 QSIs. Probability of unlicensed CC availability $P_{off}$ is Gaussian with mean value of $\mu=0.5$ and standard deviation $\phi^2=0.1$.

Our intuition is validated with the help Figure \ref{quality_cdf} since CDFs with more UEs are higher than those with fewer UEs, meaning that they end up with more low quality segments. More specifically for $K=15$, more than 90\% of segments are delivered in the 3 lowest quality levels of 1, 2.5 and 5 Mbps and the rest of them in just 8 Mbps. On the other hand, for $K=3$, very few segments are delivered in the 4 lowest qualities, while most of them are delivered in 10 Mbps and a small portion of under 10\% is even delivered in the highest quality of 35 Mbps.

\begin{figure}
	\begin{center}
		\includegraphics[keepaspectratio,width = 0.9\linewidth]{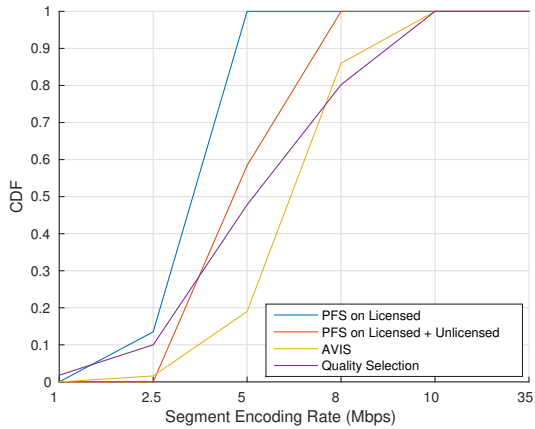}
		\caption{Segment quality CDF for PFS, AVIS and Quality Selection algorithm.}
		\label{quality_cdf_2}
	\end{center}
\end{figure}

In addition to the proposed Algorithms \ref{Algo_1} and \ref{Algo_2}, a standard PFS solution, as well as the AVIS framework are also considered for comparison. PFS strives to provide approximately the same QoS to all UEs. PFS has no knowledge about the unlicensed CC availability to the UEs that request the segments and this affects the CDF. On the other hand, since AVIS quality selection is implemented on the eNodeB, there is knowledge about the number of available resources but the scheduling is handled similarly to PFS. Figure \ref{quality_cdf_2} displays the average segment quality CDF for PFS only on a licensed CC, on a licensed plus unlicensed CC through CA, AVIS, as well as Algorithm \ref{Algo_1} solution. The results were obtained for a number of $K=10$ UEs and a duration of 100 QSIs.

First, by comparing the two PFS results we can see how the utilization of a secondary CC in the unlicensed band has increased average video quality. All users are served by segments the encoding rate of which does not exceed 5 Mbps on a licensed only system. However, when an unlicensed CC is added, a percentage of about 60\% is served by the 3 lowest qualities (1,2.5,5 Mbps) while the remaining users experience quality of 8 Mbps segments. However, when Algorithm 1 is used to select segment quality we can see that more UEs are served with segment encoding rates above 5 Mbps compared to the PFS solutions. There is however a small 10\% of UEs where the two lowest qualities are chosen while in the respective case of PFS this percentage is 0. In contrast to PFS, Algorithm 1 allocates more resources to UEs with good channel conditions and fewer (thus the worse quality) to the ones experiencing bad channel conditions. In anycase the percentage of UEs with high segment quality is significantly higher than the ones of low segment quality highlighting the usefulness of Algorithm 1. Comparing AVIS to Algorithm \ref{Algo_1} however we can see that only about 20\% of segments are delivered in the 3 lowest quality levels while for most of them, the 8 Mbps encoding rate is chosen. This indicates a better performance than Algorithm \ref{Algo_1} but we observe once again that Algorithm \ref{Algo_1} leads to a larger variety of encoding rates providing more 10 Mbps segments than AVIS. AVIS does not consider UE buffer status and thus never decides to assign a lower quality if the previous segment could not be delivered, thus making the entire framework less adaptive but more aggressive, impacting this way buffer under-run occurrences.

\subsubsection{Video freezes}
\begin{figure}
	\begin{center}
		\includegraphics[keepaspectratio,width = 0.9\linewidth]{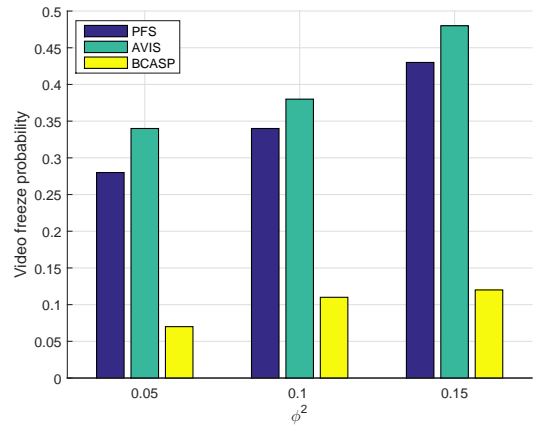}
		\caption{Video freeze probability comparison between PFS, AVIS and BCASP.}
		\label{freeze_prob}
	\end{center}
\end{figure}

The user's QoE is not only determined by the quality of the video displayed. Playback should be smooth with minimum to none interruptions for buffering to ensure maximum viewing experience. BCASP was designed so that the selected video segments will be delivered to the UEs before their playback is due. What makes on time delivery challenging in our system is the fact the unlicensed CC access probability varies over time, and the system cannot provide stable data rates to the UEs. Thus, Figure \ref{freeze_prob} displays the average freeze probability for an LAA system where the mean value of $P_{off}$ is $\mu=0.5$ and its standard deviation takes values from the set [0.05,0.1,0.15]. As the standard deviation increases, the freeze probability is also expected to increase, since unlicensed CC availability varies more through QSIs and so does the average rate, increasing the chance of selecting video qualities that cannot be delivered in the following QSI.

In all tested cases of Figure \ref{freeze_prob} the dominance of BCASP over PFS and AVIS is evident since the video freeze probability is about 3 times lower with BCASP. This is because the eNodeB decides segment quality by considering unlicensed CC traffic dynamics. This helps in making better decisions in contrast to PFS where resources of the unlicensed CC are scheduled upon being available resulting in fluctuating data rates and thus mistaken segment quality selections by the UEs that are unable to see the bigger picture. As for the AVIS framework, the aggressive policy of selecting high quality segments impacts freeze probability since the scheduler (PFS) takes no action in order to empty the transmission queues which are anyway difficult to empty due to the quality selection decisions. This results in an even higher freeze probability than PFS.

\begin{figure}
	\begin{center}
		\includegraphics[keepaspectratio,width = 0.9\linewidth]{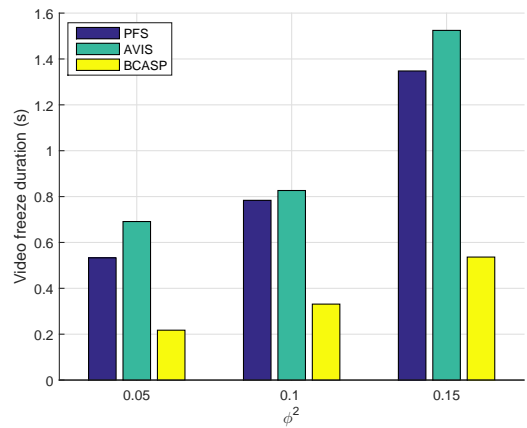}
		\caption{Video freeze duration comparison between PFS, AVIS and BCASP.}
		\label{freeze_dur}
	\end{center}
\end{figure}

The impact of video freezes on QoE depends not only in their frequency of occurrence but also in their duration. A short freeze duration of a couple hundred of milliseconds might even be negligible compared to one that lasts a couple of seconds and maybe more. In Figure \ref{freeze_dur} we once again compare PFS, AVIS and BCASP in terms of average freeze duration in seconds for the same $P_{off}$ standard deviation values tested in Figure \ref{freeze_prob}.

Once again we observe the superiority of BCASP compared to PFS and AVIS. BCASP schedules resources by considering the queue lengths of the UEs and tries to empty the longest ones. This implies that even in the event of a video freeze and since the scheduler has made its best effort to empty the queues, the remaining duration of the segments not yet delivered is kept at minimum. With PFS on the other hand, there is no such guarantee and the buffering duration is increased. Once again we highlight that the reason why BCASP performs so much better concerning video freezes is that on one hand, the segment qualities have been selected by the eNodeB which has unlicensed CC access probability knowledge, and on the other hand BCASP considers backlogs. AVIS again shows longer freeze duration than both PFS and BCASP since queue backlogs are large and difficult to empty with a PFS approach.

\section{Conclusion}
In this work we presented a framework for the application of adaptive video streaming over an LTE Unlicensed system based on LAA. Unlicensed channel access is a key enabler for 5G radio communications towards increasing data rates and satisfying demanding applications such as adaptive video streaming. To this end, the presented analysis exploits the extra unlicensed spectrum by placing segment quality decisions to the eNodeB, which has unlicensed band activity knowledge. Quality selection is accomplished by employing a utility maximization problem at the eNodeB under resource allocation constraints. The convex problem is solved using ADMM and its solution determines the segment qualities to be delivered to the users for the next interval. Furthermore, a scheduling policy that is ideal for delivering the predetermined amount of payload under specific time constraints is employed. The policy is based on Lyapunov optimization and its advantage is that schedules resources not only to the users that experience high instant data rate, but also have increased queue backlogs, meaning that there are data that must be transmitted for completing the segment download. Our scheme is compared with a standard PFS approach and a state-of-the-art LTE-A adaptive streaming system.

\bibliography{citations}

\end{document}